\documentclass[fleqn,twoside]{article}
\usepackage[headings]{espcrc2}

\usepackage{amsmath}
\usepackage{amsfonts}
\usepackage{cite}

\readRCS
$Id: espcrc2.tex,v 1.2 2004/02/24 11:22:11 spepping Exp $
\usepackage{graphicx}
\usepackage[figuresright]{rotating}

\def\bq{\begin{eqnarray}}
\def\eq{\end{eqnarray}}

\def\eps{\varepsilon}

\title{Blowing up Feynman integrals}

\author{Christian Bogner$^a$ and Stefan Weinzierl\address{Institut f{\"u}r Physik, Universit{\"a}t Mainz,
        D - 55099 Mainz, Germany}
       }
       
\runtitle{Blowing up Feynman integrals}
\runauthor{Ch. Bogner, S. Weinzierl}

\begin{document}

\begin{abstract}
In this talk we discuss sector decomposition.
This is a method to disentangle overlapping singularities
through a sequence of blow-ups.
We report on an open-source implementation of this algorithm
to compute numerically the Laurent expansion
of divergent multi-loop integrals.
We also show how this method can be used to prove a theorem 
which relates the coefficients of the Laurent series
of dimensionally regulated multi-loop integrals to periods.
\vspace{1pc}
\end{abstract}

\maketitle

% -----------------------------------------------------------------------
\section{INTRODUCTION}

The calculation of higher-order corrections in perturbation theory in particle physics
relies to a large extent on our abilities to compute Feynman loop integrals.
These calculations are complicated by the occurrence 
of ultraviolet and infrared singularities.
Ultraviolet divergences are related to the high-energy behaviour of the integrand.
Infrared divergences may occur if massless particles are present in the theory
and are related to the low-energy or collinear behaviour of the integrand.

Dimensional regularisation
is usually employed to regularise these singularities.
Within dimensional regularisation one considers the loop integral in $D$ space-time dimensions
instead of the usual four space-time dimensions.
The result is expanded as a Laurent series in the parameter $\eps=(4-D)/2$, describing the deviation
of the $D$-dimensional space from the usual four-dimensional space.
The singularities manifest themselves as poles in $1/\eps$.
Each loop can contribute a factor $1/\eps$ from the ultraviolet divergence and a factor $1/\eps^2$
from the infrared divergences. 
Therefore an integral corresponding to a graph with $l$ loops can have poles up to $1/\eps^{2l}$.

At the end of the day, all poles disappear: The poles related to ultraviolet divergences
are absorbed into renormalisation constants.
The poles related to infrared divergences cancel in the final result for infrared-safe observables, 
when summed over all degenerate states 
or are absorbed into universal parton distribution functions.
The sum over all degenerate states involves a sum over contributions with different
loop numbers and different numbers of external legs.

However, intermediate results are in general a Laurent series in $\eps$
and the task is to determine the coefficients of this Laurent series up to a certain order.
Sector decomposition \cite{Hepp:1966eg,Roth:1996pd,Binoth:2000ps,Binoth:2003ak,Bogner:2007cr,Heinrich:2008si}
can be used to compute numerically the coefficients of the Laurent series. 
At the same time, sector decomposition provides the tools to prove that under rather weak assumptions
the coefficients of the Laurent series are special numbers called periods \cite{Bogner:2007mn}.

% -----------------------------------------------------------------------
\section{MULTI-LOOP INTEGRALS}

Let us now consider a generic scalar $l$-loop integral $I_G$ 
in $D=2m-2\eps$ dimensions with $n$ propagators,
corresponding to a graph $G$.
Let us further make a slight generalisation: 
For each internal line $j$ the corresponding propagator
in the integrand can be raised to a power $\nu_j$.
Therefore the integral will depend also on the numbers $\nu_1$,...,$\nu_n$.
It is sufficient to consider only the case, where all exponents are natural numbers: $\nu_j \in {\mathbb N}$.
We define the Feynman integral by
\bq
\label{eq0}
\lefteqn{
I_G  = 
 \frac{\prod\limits_{j=1}^{n}\Gamma(\nu_j)}{\Gamma(\nu-lD/2)}
 \left( \mu^2 \right)^{\nu-l D/2}
} & & \nonumber \\
 & &
 \times
 \int \prod\limits_{r=1}^{l} \frac{d^Dk_r}{i\pi^{\frac{D}{2}}}\;
 \prod\limits_{j=1}^{n} \frac{1}{(-q_j^2+m_j^2)^{\nu_j}},
\eq
with $\nu=\nu_1+...+\nu_n$.
The momenta $q_j$ of 
the propagators are linear combinations of the external momenta and the loop
momenta.
The prefactors are chosen such that after Feynman parameterisation the Feynman integral
has a simple form:
\bq
\label{eq1}
\lefteqn{
I_G  = 
 \left( \mu^2 \right)^{\nu-l D/2}
 \int\limits_{x_j \ge 0}  d^nx \;
 \delta(1-\sum_{i=1}^n x_i)\,
} & & \nonumber \\
 & &
 \times
 \left( \prod\limits_{j=1}^n x_j^{\nu_j-1} \right)
 \frac{{\mathcal U}^{\nu-(l+1) D/2}}{{\mathcal F}^{\nu-l D/2}}.
\eq
The functions ${\mathcal U}$ and $\mathcal F$ depend on the Feynman parameters
and can be derived
from the topology of the corresponding Feynman graph $G$.
Cutting $l$ lines of a given connected $l$-loop graph such that it becomes a connected
tree graph $T$ defines a chord ${\mathcal C}(T,G)$ as being the set of lines 
not belonging to this tree. The Feynman parameters associated with each chord 
define a monomial of degree $l$. The set of all such trees (or 1-trees) 
is denoted by ${\mathcal T}_1$.  The 1-trees $T \in {\mathcal T}_1$ define 
${\mathcal U}$ as being the sum over all monomials corresponding 
to the chords ${\mathcal C}(T,G)$.
Cutting one more line of a 1-tree leads to two disconnected trees $(T_1,T_2)$, or a 2-tree.
${\mathcal T}_2$ is the set of all such  pairs.
The corresponding chords define  monomials of degree $l+1$. Each 2-tree of a graph
corresponds to a cut defined by cutting the lines which connected the two now disconnected trees
in the original graph. 
The square of the sum of momenta through the cut lines 
of one of the two disconnected trees $T_1$ or $T_2$
defines a Lorentz invariant
\bq
s_{T} & = & \left( \sum\limits_{j\in {\mathcal C}(T,G)} p_j \right)^2.
\eq   
The function ${\mathcal F}_0$ is the sum over all such monomials times 
minus the corresponding invariant. The function ${\mathcal F}$ is then given by ${\mathcal F}_0$ plus an additional piece
involving the internal masses $m_j$.
In summary, the functions ${\mathcal U}$ and ${\mathcal F}$ are obtained from the graph as follows:
\bq
\label{eq0def}	
 {\mathcal U} 
 & = & 
 \sum\limits_{T\in {\mathcal T}_1} \Bigl[\prod\limits_{j\in {\mathcal C}(T,G)}x_j\Bigr]\;,
 \nonumber\\
 {\mathcal F}_0 
 & = & 
 \sum\limits_{(T_1,T_2)\in {\mathcal T}_2}\;\Bigl[ \prod\limits_{j\in {\mathcal C}(T_1,G)} x_j \Bigr]\, (-s_{T_1})\;,
 \nonumber\\
 {\mathcal F} 
 & = &  
 {\mathcal F}_0 + {\mathcal U} \sum\limits_{j=1}^{n} x_j m_j^2\;.
\eq
In eq.~(\ref{eq0}) we just considered scalar integrals.
A priori more complicated cases, where the loop momentum appears in the numerator might occur.
However, there is a general reduction algorithm, which reduces these tensor integrals
to scalar integrals \cite{Tarasov:1996br,Tarasov:1997kx}.
The price we have to pay is that these scalar integrals involve higher powers of the propagators
and/or have dimensions shifted by two units.
Therefore we introduced in eq.~(\ref{eq0}) higher powers of the propagators
and kept the dimension $D=2m-2\eps$ arbitrary.
As a consequence, integrals of the form as in eq.~(\ref{eq0}) 
are the most general loop integrals we have to consider.

% -----------------------------------------------------------------------
\section{SECTOR DECOMPOSITION}

In this section we review the algorithm for iterated
sector decomposition.
The starting point is an integral of the form
\bq
\label{basic_integral}
\lefteqn{
\hspace*{-8mm}
 \int\limits_{x_j \ge 0} d^nx \;\delta(1-\sum_{i=1}^n x_i)
 \left( \prod\limits_{i=1}^n x_i^{\mu_i} \right)
 \prod\limits_{j=1}^r \left[ P_j(x) \right]^{\lambda_j},
} & &
 \nonumber \\
\eq
where $\mu_i=a_i+\eps b_i$ and $\lambda_j=c_j+\eps d_j$.
The integration is over the standard simplex.
The $a$'s, $b$'s, $c$'s and $d$'s are integers.
The $P$'s are polynomials in the variables $x_1$, ..., $x_n$.
The polynomials are required to be non-zero
inside the integration region, but
may vanish on the boundaries of the integration region.
The algorithm consists of the following steps:
\\
\\
Step 0: Convert all polynomials to homogeneous polynomials.
\\
\\
Step 1: Decompose the integral into $n$ primary sectors.
\\
\\
Step 2: Decompose the sectors iteratively into sub-sectors until each of the polynomials 
is of the form
\bq
\label{monomialised}
 P & = & x_1^{m_1} ... x_n^{m_n} \left( c + P'(x) \right),
\eq
where $c\neq 0$ and $P'(x)$ is a polynomial in the variables $x_j$ without a constant term.
In this case the monomial prefactor $x_1^{m_1} ... x_n^{m_n}$ can be factored out
and the remainder contains a non-zero constant term.
To convert $P$ into the form~(\ref{monomialised}) one chooses a subset
$S=\left\{ \alpha_{1},\,...,\, \alpha_{k}\right\} \subseteq \left\{ 1, \,...\, n \right\}$
according to a strategy discussed in the next section.
One decomposes the $k$-dimensional hypercube into $k$ sub-sectors according to
\bq
\label{decomposition}
\hspace*{-7mm}
 \int\limits_{0}^{1} d^{n}x & = & 
 \sum\limits_{l=1}^{k} 
 \int\limits_{0}^{1} d^{n}x
   \prod\limits_{i=1, i\neq l}^{k}
   \theta\left(x_{\alpha_{l}}\geq x_{\alpha_{i}}\right).
\eq
In the $l$-th sub-sector one makes for each element of $S$ the
substitution
\bq
\label{substitution}
x_{\alpha_{i}} & = & x_{\alpha_{l}} x_{\alpha_{i}}' \;\;\;\mbox{for}\; i\neq l.
\eq
This procedure is iterated, until all polynomials are of the form~(\ref{monomialised}).
\begin{figure}[t]
\includegraphics[bb= 100 480 570 710,width=0.45\textwidth]{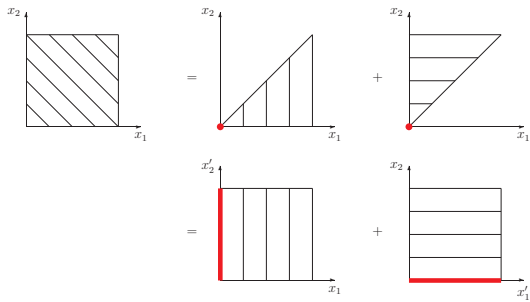}
\caption{\label{fig1} Illustration of sector decomposition and blow-up for a simple example.}
\end{figure}
Fig.~\ref{fig1} illustrates this for the simple example $S=\{1,2\}$. 
Eq.~(\ref{decomposition}) gives the decomposition into the two sectors
$x_1>x_2$ and $x_2>x_1$.
Eq.~(\ref{substitution}) transforms the triangles into squares.
This transformation is one-to-one for all points except
the origin. The origin is replaced by the line $x_1=0$ in the first sector
and by the line $x_2=0$ in the second sector.
Therefore the name ``blow-up''.
\\
\\
Step 3: The singular behaviour of the integral depends now only on the factor
\bq
  \prod\limits_{i=1}^{n}x_{i}^{a_{i}+\epsilon b_{i}}.
\eq
We Taylor expand in the integration variables and perform the trivial integrations
\bq
 \int\limits_{0}^{1} dx \; x^{a+b\eps}
 & = &
 \frac{1}{a+1+b\eps},
\eq
leading to the explicit poles in $1/\eps$.
\\
\\
Step 4: All remaining integrals are now by construction finite.
We can now expand all expressions in a Laurent series in $\eps$
and truncate to the desired order.
\\
\\
Step 5: It remains to compute the coefficients of the Laurent series.
These coefficients contain finite integrals, which can be evaluated numerically
by Monte Carlo integration.

% -----------------------------------------------------------------------
\section{HIRONAKA'S POLYHEDRA GAME}

In step 2 of the algorithm we have an iteration.
It is important to show that this iteration terminates and does not lead to an infinite
loop.
\begin{figure*}[t]
\includegraphics[bb= 100 600 700 710,width=\textwidth]{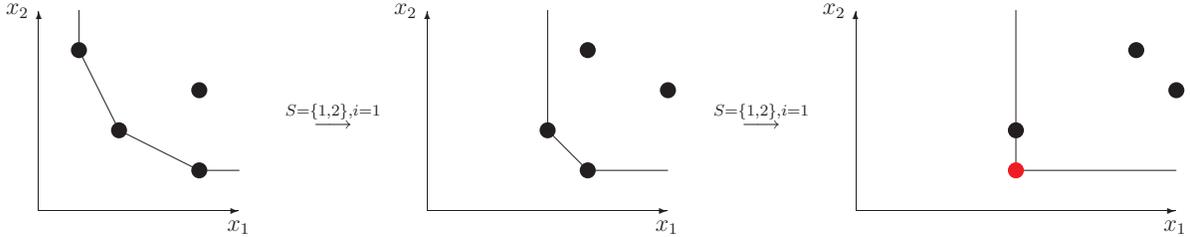}
\caption{\label{fig2} Illustration of Hironaka's polyhedra game.}
\end{figure*}
There are strategies for choosing the sub-sectors, which guarantee termination.
These strategies \cite{Hironaka:1964,Spivakovsky:1983,Encinas:2002,Hauser:2003,Zeillinger:2006} are closely related to Hironaka's polyhedra game.

Hironaka's polyhedra game is played by two players, A and B. They are
given a finite set $M$ of points $m=\left(m_{1},\,...,\,m_{n}\right)$
in $\mathbb{N}_{+}^{n}$, the first quadrant of $\mathbb{N}^{n}$.
We denote by $\Delta \subset\mathbb{R}_{+}^{n}$ the positive convex hull of the set $M$.
It is given by the convex hull of the set
\bq
\bigcup\limits_{m\in M}\left(m+\mathbb{R}_{+}^{n}\right).
\eq
The two players compete in the following game:
\begin{enumerate}
\item Player A chooses a non-empty subset $S\subseteq\left\{ 1,\,...,\, n\right\}$.
\item Player B chooses one element $i$ out of this subset $S$. 
\end{enumerate}
Then, according
to the choices of the players, the components of all $\left(m_{1},\,...,\,m_{n}\right)\in M$
are replaced by new points $\left(m_{1}^{\prime},\,...,\,m_{n}^{\prime}\right)$,
given by:
\bq
\label{update_polyhedron}
m_{j}^{\prime} & = & m_{j}, \;\;\; \textrm{if }j\neq i, \nonumber \\
m_{i}^{\prime} & = & \sum_{j\in S} m_{j}-c,
\eq
where for the moment we set $c=1$. 
This defines the set $M^\prime$.
One then sets $M=M^\prime$ and goes back to step 1.
Player A wins the game if, after a finite number of moves, 
the polyhedron $\Delta$ is of the form
\bq
\label{termination}
 \Delta & = & m+\mathbb{R}_{+}^{n},
\eq
i.e. generated by one point.
If this never occurs, player $B$ has won.
The challenge of the polyhedra game is to show that player $A$ always has
a winning strategy, no matter how player $B$ chooses his moves.
A simple illustration of Hironaka's polyhedra game in two dimensions is given in
fig.~\ref{fig2}. Player A always chooses $S=\{1,2\}$.
In ref.~\cite{Bogner:2007cr} we have shown that a winning strategy for 
Hironaka's polyhedra game
translates directly into a strategy for choosing the sub-sectors which
guarantees termination.

% -----------------------------------------------------------------------
\section{AN OPEN-SOURCE IMPLEMENTATION}

The algorithm for computing numerically the coefficients of the Laurent series
of a divergent multi-loop integral involves symbolic computer algebra and numerical
integration.
Partially due to this ``double nature'' of the algorithm, 
no public program was available until recently.
We implemented\footnote{The program can be obtained from {\tt http://www.higgs.de/\~{}stefanw/software.html}}
the algorithm in the language C++ using the library GiNaC\cite{Bauer:2000cp}
for the symbolic part.
In addition the GNU Scientific Library \cite{GSL} is used for the numerical Monte Carlo integration.
We have implemented various strategies for choosing the sub-sectors.
Three of them are guaranteed to terminate. However in practice they lead to a rather large
number of sub-sectors.
In addition we have implemented two heuristic strategies,
for which we neither have a proof that they terminate,
nor do we know a counter-example which would lead to an infinite recursion.
If these heuristic strategies terminate, the number of the generated sub-sectors
tends to be smaller than the corresponding numbers for the other strategies.

% -----------------------------------------------------------------------
\section{PERIODS}

Periods are special numbers.
Before we give the definition, let us start with some sets of numbers:
The natural numbers $\mathbb{N}$,
the integer numbers $\mathbb{Z}$,
the rational numbers $\mathbb{Q}$,
the real numbers $\mathbb{R}$ and 
the complex numbers $\mathbb{C}$
are all well-known. More refined is already the set of algebraic numbers, 
denoted by $\bar{\mathbb{Q}}$.
An algebraic number is a solution of a polynomial equation with rational
coefficients:
\bq
 x^n + a_{n-1} x^{n-1} + \cdots + a_0 = 0,
 \;\;\; a_j \in \mathbb{Q}.
\eq
As all such solutions lie in $\mathbb{C}$, the set of algebraic numbers $\bar{\mathbb{Q}}$ 
is a sub-set of
the complex numbers $\mathbb{C}$.
Numbers which are not algebraic are called transcendental.
The sets $\mathbb{N}$, $\mathbb{Z}$, $\mathbb{Q}$ and $\bar{\mathbb{Q}}$ are countable, whereas
the sets $\mathbb{R}$, $\mathbb{C}$ and the set of transcendental numbers are uncountable.

Periods are a countable set of numbers, lying between $\bar{\mathbb{Q}}$ and $\mathbb{C}$.
There are several equivalent definitions for periods.
Kontsevich and Zagier gave the following definition \cite{Kontsevich:2001}:
A period is a complex number whose real and imaginary parts are values
of absolutely convergent integrals of rational functions with rational coefficients,
over domains in $\mathbb{R}^n$ given by polynomial inequalities with rational coefficients.
Domains defined by polynomial inequalities with rational coefficients
are called semi-algebraic sets.

We denote the set of periods by $\mathbb{P}$. The algebraic numbers are contained in the set of periods:
$\bar{\mathbb{Q}} \in \mathbb{P}$.
In addition, $\mathbb{P}$ contains transcendental numbers, an example for such a number is $\pi$:
\bq
 \pi & = & \iint\limits_{x^2+y^2\le1} dx \; dy.
\eq
The integral on the r.h.s. clearly shows that $\pi$ is a period.
On the other hand, it is conjectured that the basis of the natural logarithm $e$
and Euler's constant $\gamma_E$
are not periods.
Although there are uncountably many numbers, which are not periods, only very recently an example
for a number which is not a period has been found \cite{Yoshinaga:2008}.

We need a few basic properties of periods:
The set of periods $\mathbb{P}$ is a $\bar{\mathbb{Q}}$-algebra \cite{Kontsevich:2001,Friedrich:2005}.
In particular the sum and the product of two periods are again periods.

The defining integrals of periods have integrands, which are rational
functions with rational coefficients.
For our purposes this is too restrictive, as we will encounter
logarithms as integrands as well.
However any logarithm of a rational function with rational coefficients can be written as
\bq
 \ln g(x) 
 & = &
 \int\limits_0^1 dt \; \frac{g(x)-1}{(g(x)-1) t + 1}.
\eq

% -----------------------------------------------------------------------
\section{A THEOREM ON FEYNMAN INTEGRALS}

Let us consider a general scalar multi-loop integral as in eq.~(\ref{eq1})
Let $m$ be an integer and set $D=2 m - 2 \eps$. Then this integral has 
a Laurent series expansion in $\eps$
\bq
 I_G & = & \sum\limits_{j=-2l}^\infty c_j \eps^j.
\eq
{\bf Theorem 1}: In the case where
\begin{enumerate}
\item all kinematical invariants $s_T$ are zero or negative, 
\item all masses $m_i$ and $\mu$ are zero or positive ($\mu\neq0$),
\item all ratios of invariants and masses are rational,
\end{enumerate}
the coefficients $c_j$ of the Laurent expansion are periods.

In the special case were
\begin{enumerate}
\item the graph has no external lines or all invariants $s_T$ are zero,
\item all internal masses $m_j$ are equal to $\mu$,
\item all propagators occur with power $1$, i.e. $\nu_j=1$ for all $j$,
\end{enumerate}
the Feynman parameter integral reduces to
\bq
I_G  & = &
 \int\limits_{x_j \ge 0}  d^nx \;
 \delta(1-\sum_{i=1}^n x_i)\,
 {\mathcal U}^{- D/2}
\eq
and only the polynomial ${\cal U}$ occurs in the integrand.
In this case it has been shown by Belkale and Brosnan \cite{Belkale:2003} that the coefficients of the 
Laurent expansion are periods.

Using the method of sector decomposition we are able to prove the general case.
We will actually prove a stronger version of theorem 1.
Consider the following integral
\bq
\label{basic_integral2}
\lefteqn{
J = 
 \int\limits_{x_j \ge 0} d^nx \;\delta(1-\sum_{i=1}^n x_i)
} & & \nonumber \\
 & &
 \left( \prod\limits_{i=1}^n x_i^{a_i+\eps b_i} \right)
 \prod\limits_{j=1}^r \left[ P_j(x) \right]^{d_j+\eps f_j}.
\eq
The integration is over the standard simplex.
The $a$'s, $b$'s, $d$'s and $f$'s are integers.
The $P$'s are polynomials in the variables $x_1$, ..., $x_n$ with rational coefficients.
The polynomials are required to be non-zero
inside the integration region, but
may vanish on the boundaries of the integration region.
To fix the sign, let us agree that all polynomials are positive inside the integration region.
The integral $J$ has a Laurent expansion
\bq
 J & = & \sum\limits_{j=j_0}^\infty c_j \eps^j.
\eq
{\bf Theorem 2}: The coefficients $c_j$ of the Laurent expansion of the integral $J$ are periods.

Theorem 1 follows then from theorem 2 as the special case
$a_i=\nu_i-1$, $b_i=0$, $r=2$, $P_1={\cal U}$, $P_2={\cal F}$, 
$d_1+\eps f_1 = \nu-(l+1)D/2$ and $d_2+\eps f_2 = l D/2 - \nu$.

Proof of theorem 2:
To prove the theorem we will give an algorithm which expresses each coefficient $c_j$
as a sum of absolutely convergent integrals over the unit hypercube with integrands,
which are linear combinations 
of products of rational functions with logarithms of rational functions,
all of them with rational coefficients.
Let us denote this set of functions to which the integrands belong by ${\cal M}$.

The unit hypercube is clearly a semi-algebraic set.
It is clear that absolutely convergent integrals
over semi-algebraic sets with integrands from the set ${\cal M}$ are periods.
In addition, the sum of periods is again a period.
Therefore it is sufficient to express each coefficient $c_j$ as a finite sum
of absolutely convergent integrals over the unit hypercube with integrands from ${\cal M}$.
To do so, we use iterated sector decomposition.
Details of the proof can be found in \cite{Bogner:2007mn}.

% -----------------------------------------------------------------------
\section{CONCLUSIONS}

In this talk we reported on an open-source implementation of a computer program
to compute numerically the coefficients of the Laurent expansion of divergent multi-loop
integrals.
We also showed that under rather weak assumptions all these coefficients are periods.
The technique we used was based on sector decomposition and sequences of blow-ups.
It appears that Feynman integrals and algebraic geometry are not un-related, as was already pointed out
in \cite{Bloch:2005}.

% ----------------------------------------------
% references
\bibliography{/home/stefanw/notes/biblio}
\bibliographystyle{h-elsevier2}

\end{document}